\documentclass[12pt,twoside]{article}
\usepackage{graphicx}
\usepackage{amsfonts}
\usepackage{latexsym}
\usepackage{multirow}

\newsavebox{\tallguy}
\savebox{\tallguy}{\mbox{\rule{0ex}{2.5ex}}}

\newcommand{\mket}[1]{\left | #1 \right ) }
\newcommand{\mbra}[1]{\left ( #1 \right | }
\newcommand{\mamp}[2]{\left ( #1 \left | #2 \right. \right ) }

\newcommand{\mresult}[4]{\left \langle {#1}={#2}:{#3}={#4} \right \rangle }

\newcommand{\modalspace}{\mbox{$\mathcal{V}$}}

\newcommand{\dual}[1]{{#1}^{\ast}}

\newcommand{\scalarfield}{\mbox{$\mathcal{F}$}}
\newcommand{\zeetwo}{\mbox{$\mathbb{Z}_{2}$}}

\newcommand{\sys}[1]{^{\mbox{\tiny (#1)}}}

\newcommand{\onehalf}{\mbox{$\frac{1}{2}$}}

\title{Non-contextuality and free will in modal quantum theory}
\author{Benjamin Schumacher\footnote{Department of Physics, Kenyon College.  Email schumacherb@kenyon.edu} 
    \,\, and Michael D. Westmoreland\footnote{Department of Mathematical Sciences, Denison University.  Email westmoreland@denison.edu}}\date{Kenyon College and Denison University}
\begin{document}

\maketitle
\thispagestyle{empty}

\begin{abstract}
  Modal quantum theory (MQT) is a simplified cousin of ordinary
  Hilbert space quantum theory.  We show that two important
  theorems of actual quantum theory, the Kochen-Specker
  theorem excluding non-contextual hidden variables and the
  Conway-Kochen ``free will theorem'' about entangled systems,
  have direct analogues in MQT.  The proofs of these analogue
  theorems are similar to, but much simpler than, the originals.
  We also show that the structure of possible measurement
  results for an entangled system in MQT cannot be represented
  by probability assignments satisfying the no-signaling principle,
  such as those given by ordinary quantum theory.
\end{abstract}

\section{Modal quantum theory}

Modal quantum theory, or MQT, is a simplified mathematical
model having many affinities with the Hilbert space structure
actual quantum theory (AQT).  In \cite{mqtpaper} it was shown that,
though the state space in MQT may be discrete and lacks an inner 
product, it is nevertheless possible to give a usable interpretation
of the model based on the ``modal'' distinction between {\em possible} 
and {\em impossible} measurement outcomes.  MQT systems have superposition
and interference effects, and entangled systems have properties
that are inconsistent with local hidden variable theories.

In this paper, we show that MQT also supports analogues of two
important results from actual quantum theory, namely the 
Kochen-Specker argument about contextuality and hidden variable
theories \cite{kspaper}, and the Conway-Kochen ``free will 
theorem'' about the properties of entangled systems \cite{freewill}.  
In each case, the proof of the MQT version of the theorem 
is substantially simpler.  We are not giving new proofs of
these important results in AQT, but rather deriving analogue
results in MQT, a very different theory.  To emphasize this,
we show that the predictions of MQT cannot always be emcompassed
by a probabilistic theory such as AQT.

The basic rules of MQT are easy to summarize.
\begin{description}
    \item[States.]  
        The state of a system is a non-zero vector $\mket{\psi}$ 
        in a vector space $\modalspace$ over a field $\scalarfield$ 
        of scalars.  $\scalarfield$ may be any field; many interesting
        examples arise when $\scalarfield$ is chosen to be finite.
    \item[Effets and measurements.]  
        An effect is an element $\mbra{e}$ in the dual space
        $\dual{\modalspace}$.  The effect $\mbra{e}$ is possible 
        for the state $\mket{\psi}$ if $\mamp{e}{\psi} \neq 0$ and 
        impossible if $\mamp{e}{\psi} = 0$.  A
        measurement is a basis for $\dual{\modalspace}$ whose
        elements correspond to the potential results of the 
        measurement process.
    \item[Composite systems.]  
        If two systems have state spaces $\modalspace\sys{1}$ and
        $\modalspace\sys{2}$, then the composite system has a 
        state space $\modalspace\sys{12} = \modalspace\sys{1} 
        \otimes \modalspace\sys{2}$.
    \item[Time evolution.]  
        The time evolution of the state of an isolated system 
        in MQT is described by an invertible linear operator
        $T$:
        \begin{equation}
            \mket{\psi} \longrightarrow \mket{\psi'} = T \mket{\psi}.
        \end{equation}
        For finite $\scalarfield$, this evolution cannot be
        continuous in time.
\end{description}
(The time evolution rule is included here for completeness; 
we will require only the first three for our discussion.)

Let us illustrate these rules by considering a system with 
$\dim \modalspace = \dim \dual{\modalspace} = 2$,  a ``modal bit''
or {\em mobit}.  The dual space contains at least three 
distinct non-zero vectors, which we can designate
$\mbra{a}$, $\mbra{b}$, and $\mbra{c} = \mbra{a} + \mbra{b}$.
(For the simplest case, when $\scalarfield = \zeetwo$, these
are in fact the only three non-zero vectors in $\dual{\modalspace}$.)

\begin{figure}
\begin{center}
\includegraphics[width=1.5in]{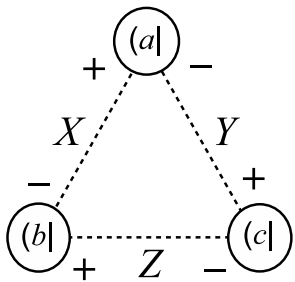}
\caption{Three measurement bases for a mobit system..
    \label{fig:mobitbases}}
\end{center}
\end{figure}

Any pair of these forms a basis for $\dual{\modalspace}$,
and thus represents a possible measurement on the system.  We shall
call the three measurement bases $X$, $Y$ and $Z$, and designate the
two potential results for each measurement by $+$ or $-$.  Thus,
\begin{equation}
    \begin{array}{clcclccl}
        X: & \mbra{+_{x}} = \mbra{a} & \quad & Y: & \mbra{+_{y}} = \mbra{c} & \quad & Z: & \mbra{+_{z}} = \mbra{b} \\
           & \mbra{-_{x}} = \mbra{b} &       &    & \mbra{-_{y}} = \mbra{a} &       &    & \mbra{-_{z}} = \mbra{c}
    \end{array} .  \label{eq:XYZbases}
\end{equation}
These are shown in Figure~\ref{fig:mobitbases}.

\section{Non-contextuality}

Kochen and Specker \cite{kspaper}
considered a single spin-1 system in AQT 
together with measurements of the squares 
of spin components along various axes.  These
measurements have three important properties:
\begin{itemize}
    \item  The possible results of any measurement are 0 and 1
        (in units of $\hbar^{2}$).
    \item  The squared spin components along any three orthogonal 
        directions are compatible observables and thus may
        be measured simultaneously.
    \item  If the squared spin components along three orthogonal
        directions are measured, then exactly two of these directions
        will yield the result 1 and one direction will yield 0.
\end{itemize}
We can view the collection of squared spin components along an
orthogonal triad of directions as a single observable, whose
result is the particular direction along which the measurement
yields 0.  Each possible result corresponds to a projection
effect operator on the spin 1 Hilbert space.  The orthogonal
triad of directions (in real space) gives rise to a set of
projections onto orthogonal one-dimensional subspaces (in 
Hilbert space).

Kochen and Specker asked whether the behavior of the spin-1
system could be explained by a hidden-variable model.  In such
a model, the result of any measurement would be predetermined
by the values of one or more hidden variables, whose underlying statistical
distribution would give rise to the the observed probabilistic
properties of the quantum system.  Kochen and Specker further
required that the hidden-variable predictions be {\em non-contextual}.
That is, for a particular effect---i.e., whether or not the squared 
spin component is 0 for a given direction---the hidden-variable
``yes/no'' prediction cannot depend on which other two
directions are included in the orthogonal triad that 
is being measured.  Could such a model of non-contextual
hidden variables account for the properties of spin-1
systems listed above?

The question can be translated into a question of graph-coloring.
Given some set of directions in real three-dimensional space,
can we color those directions red (for 1) and green (for 0) such
that any orthogonal triad of directions contains exactly two
red and one green?  Kochen and Specker showed that
this was not in general possible.  They found a set of 117
directions in space that could not be colored in this way,
since each direction was part of more than one orthogonal
triad.  (Later proofs of the theorem, such as the one given
in \cite{peres}, reduce the number of directions required, 
but all such constructions remain fairly complicated.)

We can construct an analogue of the Kochen-Specker argument
in MQT, showing that the predictions of this theory are
also incompatible with non-contextual hidden variables.
Consider the three mobit measurements listed in Equation~\ref{eq:XYZbases}
and diagrammed in Figure~\ref{fig:mobitbases}.  In a hidden-variable
model for this system, the results of all measurements are
predetermined by the hidden variables, but because the
values of these variables are unknown, more than one result
may be deemed possible in a given situation.  The requirement
of non-contextuality is that the hidden variables determine
whether or not a given effect will occur, independent of
what other effects are included in the measurement.  That is,
the hidden-variable ``yes/no'' prediction is a function only
of the effect and not of the measurement basis to which that 
effect belongs.

Therefore, our question is once again translated into a
graph-coloring problem.  Consider the triangle in 
Figure~\ref{fig:mobitbases}.  Each vertex is an effect,
and each side corresponds to a measurement basis.  We
wish to color the vertices red and green so that each
side contains exactly one green vertex.  This is obviously
impossible.  Therefore, the properties of a mobit system
in MQT cannot be explained by an model of non-contextual
hidden variables.

The essential idea is the same, both in the Kochen-Specker proof 
in AQT and in its MQT analogue.  However, the construction
in MQT is almost trivial, in sharp contrast to that of
Kochen and Specker.

\section{The free will theorem}

Conway and Kochen have used the Kochen-Specker construction to
prove the remarkable ``free will theorem'' about entangled
quantum systems \cite{freewill}, \cite{freewill2}.  
This theorem links the freedom of
observers to choose their measurements to the ``free will''
(indeterminacy) displayed by quantum systems in producing
the outcomes of those measurements.  Conway and Kochen introduce
three basic axioms for the physical world, which they designate
SPIN, TWIN and FIN.  The SPIN axiom states that spin-1 quantum
systems behave as we have described in the 
last section.  The TWIN axiom describes the correlations 
between two spin-1 systems in a ``singlet'' (total spin zero)
state.  In this entangled quantum state, measurements 
of the squared spin components along two parallel directions 
must always yield identical results.

The FIN axiom states that the speed of information transfer
has a finite limit (which might be, but is not necessarily,
the speed of light).  
\begin{figure}
\begin{center}
\includegraphics[width=4in]{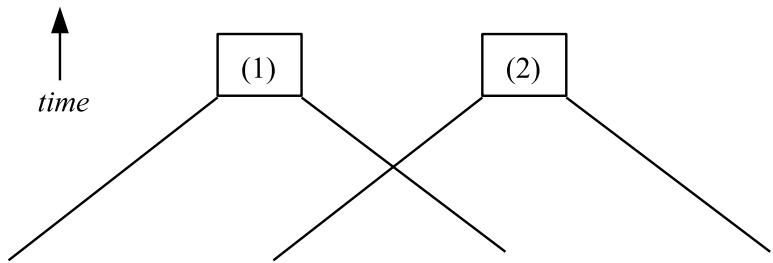}
\caption{Spacetime diagram of two near-simultaneous
    measurement processes.
    \label{fig:spacelike}}
\end{center}
\end{figure}  
Consider two measurement processes on spatially separated
subsystems, as shown in a spacetime diagram in 
Figure~\ref{fig:spacelike}.  The two measurements are 
nearly simultaneous, and so by the FIN axiom it is 
not possible for information to travel from one
measurement to the other.  This means that the result of
the measurement on system (1) can only depend on physical
conditions within the causal past of the measurement, shown
as the shaded region in Figure~\ref{fig:spacelike}.
The ``free will'' of observers means that we can arrange
the choice of system (2) measurement to lie entirely outside
the causal past of the system (1) measurement, so the system (1)
result cannot depend on this choice.  The reverse is also
true.  

These facts are the only consequences of the FIN
axiom that are used in the proof of the free will theorem.
Because of this, in a later version of the theorem Conway
and Kochen replace the FIN axiom with a weaker axiom
(called MIN) that asserts the independence of measurement
results from distant measurement choices \cite{freewill2}.

The MQT version of the free will theorem is based on the
properties of an entangled state of two mobits.  If
$\mket{0}$ and $\mket{1}$ are a pair of basis states,
this state can be written\footnote{The negative sign in
the definition of $\mket{S}$ means that the second term
is multiplied by the additive inverse of the scalar 1.
If $\scalarfield = \zeetwo$, then $-1 = 1$, and we have
$\mket{S} = \mket{0,1} + \mket{1,0}$.}
\begin{equation}
    \mket{S} = \mket{0,1} - \mket{1,0} .
\end{equation}
This state has the property that, if measurements are made
on the two mobit systems, the same effect can never occur
for both systems.  This is because, for any effect $\mbra{e}$,
\begin{equation}
    \mamp{e,e}{S} = \mamp{e}{0} \mamp{e}{1} - \mamp{e}{1} \mamp{e}{0} = 0 .
    \label{mqt-singlet}
\end{equation}
Let $\mresult{A\sys{1}}{a}{B\sys{2}}{b}$ denote the situation in which
a measurement of $A$ on system (1) yields result $a$ and a measurement
of $B$ on system (2) yields result $b$.  Then for the state $\mket{S}$,
any joint result of the form $\mresult{A\sys{1}}{a}{A\sys{2}}{a}$ 
is impossible.  The impossible results include:
\begin{equation}
    \begin{array}{ccc}
        \mresult{X\sys{1}}{+}{X\sys{2}}{+} & \quad & \mresult{X\sys{1}}{-}{X\sys{2}}{-} \\
        \mresult{Y\sys{1}}{+}{Y\sys{2}}{+} & \quad & \mresult{Y\sys{1}}{-}{Y\sys{2}}{-} \\
        \mresult{Z\sys{1}}{+}{Z\sys{2}}{+} & \quad & \mresult{Z\sys{1}}{-}{Z\sys{2}}{-}
    \end{array} .
    \label{eq:XXimpossible}
\end{equation}
Recalling the three measurement bases in Equation~\ref{eq:XYZbases},
we see that the following joint results are also impossible for $\mket{S}$:
\begin{equation}
    \begin{array}{ccc}
        \mresult{X\sys{1}}{+}{Y\sys{2}}{-} & \quad & \mresult{Y\sys{1}}{-}{X\sys{2}}{+} \\
        \mresult{Y\sys{1}}{+}{Z\sys{2}}{-} & \quad & \mresult{Z\sys{1}}{-}{Y\sys{2}}{+} \\
        \mresult{Z\sys{1}}{+}{X\sys{2}}{-} & \quad & \mresult{X\sys{1}}{-}{Z\sys{2}}{+}
    \end{array} .
    \label{eq:XYimpossible}
\end{equation}

We are now ready to state the MQT version of the three Conway-Kochen
axioms.  The FIN axiom (or its minimal replacement MIN) is retained
without change.  The MQT version of SPIN (which we may denote SPIN*)
asserts the existence of the three overlapping measurement bases 
$X$, $Y$ and $Z$ shown in Figure~\ref{fig:mobitbases}.  The MQT TWIN* axiom
states that a possible state of the system has the properties of the
MQT state $\mket{S}$---that is, that the joint results listed in 
Equations~\ref{eq:XXimpossible} and \ref{eq:XYimpossible} are impossible.

Now we are ready to prove our theorem.  Assume that FIN (or MIN), SPIN*
and TWIN* hold for a composite system.  Suppose also that observers are 
freely able to choose $X$, $Y$ and $Z$ measurements on the separated
subsystems.  Now imagine that the actual results of all measurements
are fully determined by physical factors in the causal pasts of those 
measurements.  

If the measurements $Z\sys{1}$ and $Z\sys{2}$ are made, then the results 
cannot agree.  The only two possible joint results are 
$\mresult{Z\sys{1}}{+}{Z\sys{2}}{-}$ or $\mresult{Z\sys{1}}{-}{Z\sys{2}}{+}$.
Without loss of generality (since the two situations are symmetric),
suppose the actual result is the first of these.

If the measurement on system (2) is $X\sys{2}$ instead, then the result
of the $Z\sys{1}$ measurement is unchanged, and we have
$\mresult{Z\sys{1}}{+}{X\sys{2}}{x}$ for some $x$.  According to
Equation~\ref{eq:XYimpossible}, the only possible joint result
is $\mresult{Z\sys{1}}{+}{X\sys{2}}{+}$.  We can therefore
conclude that $X\sys{2} = +$ for any choice of measurement on 
system (1).

Given that $\mresult{Z\sys{1}}{+}{Z\sys{2}}{-}$, what if the measurement
on system (1) is actually $Y\sys{1}$?  The $Z\sys{2}$ result is unchanged,
and Equation~\ref{eq:XYimpossible} tells us that only one $Y\sys{1}$
result is possible.  We must have $\mresult{Y\sys{1}}{-}{Z\sys{2}}{-}$.
From here, we can inquire how things change if we alter the system (2)
measurement to $X\sys{2}$.  Again, the $Y\sys{1}$ result cannot change,
and Equation~\ref{eq:XYimpossible} restricts us to the single
possible joint result $\mresult{Y\sys{1}}{-}{X\sys{2}}{-}$.  We can therefore
conclude that $X\sys{2} = -$ for any choice of measurement on
system (1).  This contradicts our previous conclusion about $X\sys{2}$.

Our hypothesis that the actual results of the measurements are predetermined
is therefore faulty.  Given the axioms FIN (or MIN), SPIN* and TWIN*,
the freedom of the observers to perform $X$, $Y$ or $Z$
measurements on the mobits implies that the results 
of those measurements are not
predetermined by the causal pasts of the systems.  If observers in MQT
have ``free will'', then so do the systems they observe.

The proof of the MQT version of the free will theorem has 
the same structure as the corresponding
proof by Conway and Kochen in AQT.  As in the Kochen-Specker theorem, 
the mathematical construction involved for the MQT version 
is very much simpler.

\section{Modal and actual quantum theories}

We wish to emphasize that the proofs presented here are {\em not} 
alternate proofs for the Kochen-Specker and Conway-Kochen theorems
in actual quantum theory.  These are proofs of analogous results
in modal quantum theory.  Though MQT has many affinities to AQT, 
the two are not the same.

It is conceivable, however, that systems in one theory might 
be able to simulate the other.  For example, it might be that
the possibility relations among various measurements in an MQT
system could be realized as probabilistic relations among 
measurements in an appropriately chosen AQT system.  Then the
MQT proof would apply to the corresponding AQT situation.
However, as we will now show, this cannot always be done.
In fact, the possibility relations for an MQT system need 
not be consistent with {\em any} reasonable assignment 
of probabilities for the system, much less one derived from AQT.

\begin{figure}
\begin{center}
\begin{tabular}{cc|cc|cc|cc|}
  &  & \multicolumn{2}{c}{$X\sys{2}$} & \multicolumn{2}{c}{$Y\sys{2}$} & \multicolumn{2}{c}{$Z\sys{2}$} \\
  &  & $+$ & $-$ & $+$ & $-$ & $+$ & $-$  \\ \hline 
  \multirow{2}{*}{$X\sys{1}$} & $+$ & 0 & \#    & \# & 0    & \# & \# \\
      & $-$                         & \# & 0    & \# & \#   & 0 & \# \\ \hline
  \multirow{2}{*}{$Y\sys{1}$} & $+$ & \# & \#   & 0 & \#    & \# & 0 \\
      & $-$                         & 0 & \#    & \# & 0    & \# & \# \\ \hline
  \multirow{2}{*}{$Z\sys{1}$} & $+$ & \# & 0    & \# & \#   & 0 & \# \\
      & $-$                         & \# & \#   & 0 & \#    & \# & 0 \\ \hline
\end{tabular}
\caption{Table of joint measurement results for a pair of mobits
    in the $\mket{S}$ state.  In this table, impossible results are
    designated by 0 and possible ones by \#.
    \label{fig:possibletable}}
\end{center}
\end{figure} 

Consider a pair of mobits in the entangled MQT state 
$\mket{S} = \mket{0,1}-\mket{1,0}$.  We summarize the 
predictions of this state for various joint measurements
in the table shown in Figure~\ref{fig:possibletable}.
This table contains three rows and three columns corresponding
to the $X$, $Y$ and $Z$ measurements on each mobit.  For
each joint measurement, we obtain a $2 \times 2$ sub-table
showing whether each joint result is possible (designated by \#)
or impossible (designated by 0).  The 0's in Figure~\ref{fig:possibletable}
are exactly those impossible results enumerated in
Equations~\ref{eq:XXimpossible} and \ref{eq:XYimpossible}.

We now wish to create a table of probabilities with the same essential
structure as Figure~\ref{fig:possibletable}.  The entries of the
table will be of the form $P(a,b|A\sys{1},B\sys{2})$, the probability
that a joint measurement of the observable $(A\sys{1},B\sys{2})$ will
yield the joint result $(a,b)$.  Such a table should meet the following
requirements
\begin{enumerate}
    \item  Each probability must be between 0 and 1.  Furthermore,
        for any $A\sys{1}$ and $B\sys{2}$,
        \begin{equation}
            \sum_{a,b} P(a,b|A\sys{1},B\sys{2}) = 1 .
        \end{equation}
    \item  The marginal probability distribution for the results of
        a subsystem measurement is independent of the choice of 
        measurement on the other subsystem.  For instance, given
        measurements $A\sys{1}$, ${A\sys{1}}'$ and $B\sys{2}$,
        \begin{equation}
            \sum_{a} P(a,b|A\sys{1},B\sys{2}) =
            \sum_{a'} P(a',b|{A\sys{1}}',B\sys{2}) =
            P(b|B\sys{2}) .
        \end{equation}
        This is called the {\em no-signaling principle} \cite{operational}.  
        If it were
        not true, then a choice of measurement on system (1) could cause
        an immediate change in the statistical properties of system (2), 
        allowing information to be transmitted from one to the other
        instantaneously.  The no-signaling principle holds for measurements 
        on composite systems in actual quantum theory.
    \item  Each impossible joint outcome in Figure~\ref{fig:possibletable}
        is assigned a probability \linebreak[3] $P(a,b|A\sys{1},B\sys{2}) = 0$.  
    \item  Each possible joint outcome in Figure~\ref{fig:possibletable}
        is assigned a probability \linebreak[4] $P(a,b|A\sys{1},B\sys{2}) > 0$.
\end{enumerate}
Requirements I--III almost completely specify the probability assignment
for Figure~\ref{fig:possibletable}.  The general result is shown in
Figure ~\ref{fig:probabletable}.  Only three real parameters, denoted by
$q$, $r$ and $s$ in Figure~\ref{fig:probabletable}, determine all of the
probabilities in the table.
\begin{figure}
\begin{center}
\begin{tabular}{cc|cc|cc|cc|}
  &  & \multicolumn{2}{c}{$X\sys{2}$} & \multicolumn{2}{c}{$Y\sys{2}$} & \multicolumn{2}{c}{$Z\sys{2}$} \\
  &  & $+$ & $-$ & $+$ & $-$ & $+$ & $-$  \\[1ex] \hline 
  \multirow{2}{*}{$X\sys{1}$} & $+$ & 0 & $\onehalf+q$    & $\onehalf+q$ & 0    & $\onehalf-s$ & $q+s$ \usebox{\tallguy} \\[1ex]
                              & $-$ & $\onehalf-q$ & 0    & $-q-r$ & $\onehalf+r$   & 0 & $\onehalf-q$ \\[1ex] \hline
  \multirow{2}{*}{$Y\sys{1}$} & $+$ & $\onehalf-q$ & $q+r$   & 0 & $\onehalf+r$    & $\onehalf+r$ & 0 \usebox{\tallguy}\\[1ex]
                              & $-$ & 0 & $\onehalf-r$    & $\onehalf-r$ & 0    & $-r-s$ & $\onehalf+s$ \\[1ex] \hline
  \multirow{2}{*}{$Z\sys{1}$} & $+$ & $\onehalf+s$ & 0    & $\onehalf-r$ & $r+s$   & 0 & $\onehalf+s$ \usebox{\tallguy}\\[1ex]
                              & $-$ & $-q-s$ & $\onehalf+q$   & 0 & $\onehalf-s$    & $\onehalf-s$ & 0 \\[1ex] \hline
\end{tabular}
\caption{Probability table consistent with the possibility data given
    in Figure~\ref{fig:possibletable}, consistent with the no-signaling 
    principle.  All of the entries are determined by just three
    real parameters $q$, $r$ and $s$.
    \label{fig:probabletable}}
\end{center}
\end{figure} 

From the table, however, we note that
\begin{eqnarray}
    q + r & = & P(+,-|Y\sys{1},X\sys{2}) = - P(-,+|X\sys{1},Y\sys{2}) \nonumber \\
    q + s & = & P(+,-|X\sys{1},Z\sys{2}) = - P(-,+|Z\sys{1},X\sys{2}) \\
    r + s & = & P(+,-|Z\sys{1},Y\sys{2}) = - P(-,+|Y\sys{1},Y\sys{2}) . \nonumber
\end{eqnarray}
This can only be satisfied if these six probabilities, which correspond
to possbile results in Figure~\ref{fig:possibletable}, are assigned
probability zero, in violation of requirement IV.  Therefore, 
we cannot find a probability assignment with the same structure 
as Figure~\ref{fig:possibletable} that satisfies all four requirements.  
The predictions of MQT in this case cannot be simulated by AQT, 
or indeed by any probabilistic theory satisfying the no-signaling principle.

What if we relax the troublesome requirement IV and permit an assignment
of probability zero to a result designated ``possible'' in MQT?  Then
the probability table can be completed by letting $q=r=s=0$, yielding
the table shown in Figure~\ref{fig:prcasetable}.  
It is interesting to note that, with the selection of two measurements
for each system ($X\sys{1}$, $Y\sys{1}$ for system (1) and $X\sys{2}$, $Z\sys{2}$
for system (2), for example), the probabilities from Figure~\ref{fig:prcasetable}
describe a {\em PR box}, a type of nonlocal correlation known to be 
inconsistent with AQT \cite{prbox}, \cite{operational}.  
This is also shown in Figure~\ref{fig:prcasetable}.
\begin{figure}
\begin{center}
\begin{tabular}{cc|cc|cc|cc|}
  &  & \multicolumn{2}{c}{$X\sys{2}$} & \multicolumn{2}{c}{$Y\sys{2}$} & \multicolumn{2}{c}{$Z\sys{2}$} \\
  &  & $+$ & $-$ & $+$ & $-$ & $+$ & $-$  \\[1ex] \hline 
  \multirow{2}{*}{$X\sys{1}$} & $+$ & 0 & $\onehalf$    & $\onehalf$ & 0    & $\onehalf$ & 0 \usebox{\tallguy} \\[1ex]
                              & $-$ & $\onehalf$ & 0    & 0 & $\onehalf$   & 0 & $\onehalf$ \\[1ex] \hline
  \multirow{2}{*}{$Y\sys{1}$} & $+$ & $\onehalf$ & 0   & 0 & $\onehalf$    & $\onehalf$ & 0 \usebox{\tallguy}\\[1ex]
                              & $-$ & 0 & $\onehalf$    & $\onehalf$ & 0    & $0$ & $\onehalf$ \\[1ex] \hline
  \multirow{2}{*}{$Z\sys{1}$} & $+$ & $\onehalf$ & 0    & $\onehalf$ & 0   & 0 & $\onehalf$ \usebox{\tallguy}\\[1ex]
                              & $-$ & 0 & $\onehalf$   & 0 & $\onehalf$    & $\onehalf$ & 0 \\[1ex] \hline
\end{tabular}
$\quad \Longrightarrow \quad$
\begin{tabular}{cc|cc|cc|}
  &  & \multicolumn{2}{c}{$X\sys{2}$} & \multicolumn{2}{c}{$Z\sys{2}$} \\
  &  & $+$ & $-$ & $+$ & $-$  \\[1ex] \hline 
  \multirow{2}{*}{$X\sys{1}$} & $+$ & 0 & $\onehalf$    & $\onehalf$ & 0 \usebox{\tallguy} \\[1ex]
                              & $-$ & $\onehalf$ & 0    & 0 & $\onehalf$ \\[1ex] \hline
  \multirow{2}{*}{$Y\sys{1}$} & $+$ & $\onehalf$ & 0    & $\onehalf$ & 0 \usebox{\tallguy}\\[1ex]
                              & $-$ & 0 & $\onehalf$    & $0$ & $\onehalf$ \\[1ex] \hline
\end{tabular}
\caption{Unique probability table consistent with Figure~\ref{fig:possibletable},
    satisfying only Requirements I--III.  Choosing only two rows and two columns,
    we arrive at the probability table of a PR box.
    \label{fig:prcasetable}}
\end{center}
\end{figure} 

This illustrates in the most definitive way that modal quantum theory
is not reducible to actual quantum theory.  On the other hand, there is
a sense in which AQT can be regarded a ``special case'' of MQT with
$\scalarfield = \mathbb{C}$.  
The inner product structure of AQT Hilbert space motivates us to consider
only normalized state vectors, orthonormal measurement bases, and 
unitary time evolution operators.  All of these are restrictions
of the larger class of states, measurements and evolution operators
permitted in MQT.  If we consider only the distinction
between possible ($p \neq 0$) and impossible ($p = 0$) measurement 
results, then any situation in AQT can be viewed from a modal
point of view.

The highly simplified structure of MQT is thus a generalization
of some aspects of AQT.  As we have seen, it is possible to use
this generalized framework to prove analogues to important
AQT results about hidden variable models.  The MQT proofs are
much easier to understand and can be used to shed light on the 
essential structure of these theorems.

The authors are grateful for several helpful discussions 
with H. Barnum, C. Fuchs, R. Spekkens and A. Wilce.  We
also acknowledge the hospitality of the Perimeter Institute 
for Theoretical Physics during September and October of 2010.

\end{document}